\documentclass[showpacs,aps,prd,nofootinbib,floatfix,amsmath,amssymb, twocolumn]{revtex4}
\usepackage{graphicx}
\usepackage[dvipsnames]{xcolor}	
\usepackage{calligra}	
\usepackage{graphicx}
\usepackage{hyperref}
\usepackage{float}
\restylefloat{table}
\usepackage{blindtext}
\usepackage[rightcaption]{sidecap}
\usepackage{epstopdf} %converting to PDF
\usepackage[toc,page]{appendix}
\usepackage{mathtools}

\newcommand{\al}			{\alpha}
\newcommand{\bt}			{\beta}

\newcommand{\dlt}			{\delta}
\newcommand{\eps}			{\epsilon}

\newcommand{\lm}			{\lambda}

\newcommand{\Gm}			{\Gamma}
\newcommand{\Dlt}			{\Delta}

\newcommand{\mc}[1]{\mathcal{ #1}}						% Defines command \mc{.} for calligraphy
\newcommand{\trm}[1]{\textrm{ #1}}						% Defines command \trm{.} for normal text, with spaces, in mathmode
\newcommand{\mbb}[1]{\mathbb{ #1}}						% Defines command \mbb{.} for blackboard bold capital letters
\newcommand{\ph}[1]{\phantom{ #1}}			  			% Defines command \mb{.} for bold fonts in mathmode

\newcommand{\wht}[1]{\widehat{ #1}}						% Defines command \wht{.} foa a wide hat
\newcommand{\ovl}[1]{\overline{ #1}}						% Defines command \ovl{.} for overline
\newcommand{\unl}[1]{\underline{ #1}}					% Defines command \unl{.} for underline

\newcommand{\ud}{\mathrm{d}} 								% Defines command \ud for a differential sign within an integral
\makeatletter
\newbox\slashbox \setbox\slashbox=\hbox{$/$}
\newbox\Slashbox \setbox\Slashbox=\hbox{\large$/$}
\def\pFMslash#1{\setbox\@tempboxa=\hbox{$#1$}
  \@tempdima=0.5\wd\slashbox \advance\@tempdima 0.5\wd\@tempboxa
  \copy\slashbox \kern-\@tempdima \box\@tempboxa}
\def\pFMSlash#1{\setbox\@tempboxa=\hbox{$#1$}
  \@tempdima=0.5\wd\Slashbox \advance\@tempdima 0.5\wd\@tempboxa
  \copy\Slashbox \kern-\@tempdima \box\@tempboxa}

\def\miss#1{\ifmmode{/\mkern-11mu #1}\else{${/\mkern-11mu #1}$}\fi}

\newcommand{\ket}[1]{\vert #1 \rangle}					% Defines command \ket{.} to write a ket
\newcommand{\bra}[1]{\langle #1 \vert}					% Defines command \bra{.} to write a bra
\newcommand{\brkt}[2]{\langle #1\vert #2\rangle}	% Defines command \brkt{.}{.} to write a braket

\newcommand{\tbf}[1]{\textbf{#1}}														% Bold in text mode

\newcommand{\reff}[1]{ Eq.~(\ref{#1})}															% Defines \reff to enclose the number of equation and not break line
\newcommand{\citte}[1]{~\cite{#1}}															% Defines \citte to cite without breaking line
\newcommand{\ie}{\emph{i.e.\,}}															% Defines \ie to abbreviate "id est" (that is) in italic font
\newcommand{\cf}{\emph{cf.\,}}															% Defines \cf to abbreviate "conferre" (compare) in italic font
\newcommand{\eg}{\emph{e.g.\,}}														% Defines \eg to abbreviate "exampli gratia" (for example) in italic font

\numberwithin{equation}{section}				% Add the section number to the equation label. Use together with amsmath packg,

\begin{document}

\makeatother

%\tightenlines
\title{The Feynman kernel of a dimensionally reduced scalar field theory}

\author{ M. A. L\'{o}pez-Osorio$ {}^{(a)}$}\email{maria.osorio@academicos.udg.mx}
\author{E. Mart\'{i}nez-Pascual$ {}^{(a)}$}\email{eric.martinez@academicos.udg.mx}
\author{G. N\'{a}poles-Ca\~{n}edo${}^{(b)}$}
\author{J. J. Toscano${}^{(b)}$}\email{jtoscano@fcfm.buap.mx}

\affiliation{$ ^{(a)} $Departamento de Ciencias Naturales y Exactas, \mbox{Centro Universitario de los Valles,~Universidad de Guadalajara,}\\
\mbox{Carretera Guadalajara-Ameca Km 45.5, CP 46000, Ameca, Jalisco, M\'exico.}\\ $^{(b)}$Facultad de Ciencias F\'{\i}sico Matem\'aticas,
Benem\'erita Universidad Aut\'onoma de Puebla, Apartado Postal
1152, Puebla, Puebla, M\'exico.\\
}

\begin{abstract}
We construct a consistent quantum field theory of a dimensionally reduced self-interacting scalar field. The Kaluza-Klein dimensional reduction on the well-known $\Phi^{4}$ scalar theory, on a certain $(4+n)$ spacetime with an arbitrary number of extra dimensions, induces a four dimensional reduced theory with scalar fields: the zeroth mode (``light'' field) and an infinite number of KK-excited modes (``heavy'' fields). This theory is quantized by Hamiltonian path integral methods. It is shown, from first principles, that non-trivial measure factors at the level of the functional measure are absent even if the whole set of heavy fields is taken into account.  Hints on the regularization and renormalization process are briefly discussed.
\end{abstract}

\pacs{11.10.Kk, 14.80.Rt,11.10.Gh, 31.15.xk}

\maketitle

\section{Introduction}
%\label{in}
The idea of extra dimensions in field theories dates back to the 1920s\citte{kal21,kle26}, since then, many related proposals, extensions and critical judgments about this idea have been gravitating in theoretical physics\citte{ein36,ker64,tra70,cho75,wit81}. In particular,  phenomenological string theory, naturally formulated in this framework, led the community to revisit the issue decades later from different perspectives. Using the braneworld scenario allows to consistently lower the typical scale of quantum gravity to TeVs by choosing the number and size of spatial extra dimensions\citte{ark98}  (see also\citte{lyk96}) in the so-called \emph{large}, in contrast with \emph{warped}\citte{ran99} or \emph{universal}\citte{app01}, extra dimensional models; the latter were particularly inspired by reference\citte{ant94}. Accessible lectures on these models are for instance\citte{kri04,che09}.

The basic idea in a universal extra dimensional (UED) model is rather simple\citte{hoo07}: the stage is a factorizable spacetime geometry with compact spatial extra dimensions, commonly an orbifold, on which a field theory is defined. The dimensional reduction takes place once the extra dimensional content of fields is harmonically expanded, under certain boundary conditions, and the extra dimensions themselves are integrated out at the action level; the resulting model is referred to as the effective or dimensionally reduced theory.

In this communication we are interested in the path integral quantization of the dimensionally reduced mo-del resulting from the self-interacting $\Phi^{4}$ theory on the $(4+n)$ spacetime $\mc{M}^{4}\times S^{1}/\mbb{Z}_{2}\times \cdots \times S^{1}/\mbb{Z}_{2} $, where $\mc{M}^{4}$ is the four dimensional Minkowski space time. We decided to use this model so as to do not obscure either the methods or their adaptability to other more convoluted field theories. 

Some of us have reported\citte{lop13} the equivalence at the classical level of a higher dimensional field theory and its dimensionally reduced counterpart; indeed, at the phase space level, canonical transformations can be read from the expansion of fields and momenta in the extra dimensions. In a quantum field theory, the corresponding transformations must be carefully handled. In Sec.~\ref{sec:Feyn}, we obtain the Hamiltonian path integral of the effective four dimensional theory, associated to a general higher dimensional scalar theory of first order defined on the aforementioned  $(4+n)$ spacetime; this is achieved by performing dimensional reduction at the level of the Feynman kernel. It is argued that in order to satisfy completeness and orthonormality relations of basic eigenkets of the quantum fields, the functional measures involved in the path integral, before and after the dimensional reduction, must be trivially related (see Eq.~\eqref{measures}). The trivial measure factor after the reduction process  ensures, according to\citte{unz85}, a consistent Kaluza-Klein compactification and no anomalies are brought by this process when the anomalies are understood as the noninvariance of the functional measure (see also\citte{fuj79,tom87}).  This observation is in fact necessary to properly establish the generating functional of the different Green's functions of the effective theory, which are presented in Sec.~\ref{sec:Green}.

In Sec.~\ref{secc:model} we briefly describe the dimensionally reduced model obtained from the self-interacting $\Phi^{r}$ theory (with $r\in\mbb{N}$) defined on $\mc{M}^{4}\times S^{1}/\mbb{Z}_{2}\times\cdots\times S^{1}/\mbb{Z}_{2}$ and particularize it to the case $r=4$. We point out that the infinite number of interactions in the dimensionally reduced model are all governed by a single universal coupling constant. By the end of this section, to the 1-loop approximation, the two and four point light particle vertex functions are written down. Here, the virtual contributions from all interactions are taken into account. 

In Sec.~\ref{secc:fr} we provide a plausible hint on a possible way to regularize and renormalize the theory using zeta function techniques combined with dimensional regularization.

%-------------------------------------------------------------------------------
%-------------------------------------------------------------------------------

\section{The Feynman kernel and real scalar field models with UED\lowercase{s}}
\label{sec:Feyn}

In this section we discuss the dimensional reduction of a general scalar field theory, originally defined on certain $(4+n)$ spacetime, at the level of the Feynman path integral. Our starting point is 
\begin{equation}\label{Xdim-action}
S:=\int_{\mc{M}^{4+n}}\ud x \,\mc{L}_{(4+n)}(\Phi,\partial\Phi)
\end{equation}
where $ \mc{L}_{(4+n)}$ depends on the real scalar field $ \Phi $ and its spacetime derivatives, such that the classical theory is described by second order equations of motion. At this point we regard $ \mc{M}^{4+n}=\mc{M}^{4}\times\mc{N}^{n} $ where $ \mc{M}^{4} $ is the Minkowski spacetime and the noncompact  $ n- $dimensio-nal manifold $ \mc{N}^{n} $ depicts extra spatial dimensions. A compact $\mathcal{N}^{n}$ will be considered later. All coordinate points in $ \mc{M}^{4+n} $ are written as $ (x,\bar{x}) $, or equivalently as $ (x^{M})=(x^{\mu},x^{\bar{\mu}}) $, where $ \mu=0,1,2,3 $ and $ \bar{\mu}=5,\ldots,4+n $. The field $ \Phi(x,\bar{x}) $ is a scalar with respect to both, the Poincar\'e group $ ISO(1,3+n) $ and its subgroup $ ISO(1,3) $.

As part of the compactification process\citte{lop13,lop14}, we impose $ \mc{N}^{n} $ to be the manifold $ S^{1}/\mbb{Z}_{2}\times\cdots\times S^{1}/\mbb{Z}_{2} $, where each $ S^{1} $ is assumed to have different radius $R_{i}$, $ i=1,\ldots,n $. The field $ \Phi $ is enforced to fulfill the following periodicity and parity conditions
\begin{subequations}\label{PandPcond}
\begin{align}
\Phi(x,\bar{x}) & =\Phi(x, \bar{x}+2\pi \boldsymbol{R})\ ,  \\
\Phi(x,\bar{x}) & =\Phi(x, -\bar{x}) \ ,
\end{align}
\end{subequations} 
where $\boldsymbol{R}=(R_{1},\ldots,R_{n})$, the latter is equivalent to imposing the Neumann boundary condition at the fixed points of the orbifold\citte{che09}; relations\eqref{PandPcond} allow  the Fourier expansion of the field itself
\begin{align}\label{Fourier}
\Phi(x,\bar{x}) & = \sqrt{\frac{1}{\prod_{i=1}^{n}(2\pi R_{i})}} {\phi}^{(\unl{0})}(x) \nonumber\\
&\ + \sqrt{\frac{2}{\prod_{i=1}^{n}(2\pi R_{i})}} \sum_{(\unl{m})}{\phi}^{(\unl{m})}(x)\cos\left(\frac{\unl{m}\bar{x}}{\boldsymbol{R}}\right)\
\end{align}
where the condensed notation $ (\unl{0})\equiv (0,\ldots,0) $ and $ (\unl{m})\equiv(\unl{m}_{1},\ldots,\unl{m}_{n}) $ has been used for the zero and excited KK modes, respectively. The juxtaposition $ \unl{m}\ovl{x}/\boldsymbol{R} $ stands for $ \frac{\unl{m}_{1}x^{5}}{R_{1}}+\cdots+\frac{\unl{m}_{n}x^{n+4}}{R_{n}} $; the summation symbol $ \sum_{(\unl{m})}  $ involves $ 2^{n}-1 $ different conventional nested sums as it is detailed in\citte{lop14}.

In the quantum realm, the Feynman kernel (see for instance\citte{pesbook,grebook}) 
\begin{equation}\label{FKernel}
\langle\Phi',t'\vert\Phi,t\rangle=\langle\Phi'\vert\exp[-i(t'-t)\wht{H}/\hbar]\Phi\rangle\ , 
\end{equation}
where $ \wht{H}$ is the Hamiltonian operator associated to \eqref{Xdim-action}, leads to the path integral of the theory. In order to make sense of this object, one should discretize the configuration space of the theory. As it is customary in a field theory, this is achieved by the discretization of the corresponding space\citte{grebook}. Specifically, we consider the discretization of a finite three dimensional spatial volume V (embeded in $ \mc{M}^{4} $) into $M$ cells, and the  compact extra dimensional sector $\mc{N}^{n} $ split into $K$ cells; so that the cells in the whole space are located at $ (\boldsymbol{x}_{\al},\ovl{\boldsymbol{x}}_{\bt}) $, where $ \al=1,\ldots,M $ and $ \bt=1,\ldots,K $. Therefore, on this grid, the configuration space is coarsely described   by a discrete set of variables $ \Phi_{\al \bt}(t):=\Phi(\boldsymbol{x}_{\al},t,\ovl{\boldsymbol{x}}_{\bt}) $. To each of these variables we assign the following Heisenberg operator (\cf \eqref{Fourier}):
\begin{align}\label{FiopFourier}
\wht{\Phi}_{\al \bt}(t)  & = \sqrt{\frac{1}{\prod_{i=1}^{n}(2\pi R_{i})}} \ \hat{\!\phi}^{(\unl{0})}(\boldsymbol{x}_{\al},t) \nonumber\\
&\ + \sqrt{\frac{2}{\prod_{i=1}^{n}(2\pi R_{i})}} \sum_{(\unl{m})}\ \hat{\!\phi}^{(\unl{m})}(\boldsymbol{x}_{\al},t)\cos\left(\frac{\unl{m}\bar{\boldsymbol{x}}_{\bt}}{R}\right)\ ,
\end{align}
where the operator aspect on the right hand side is attached to the Fourier coefficients in the expansion; each coefficient operator acts on the respective Hilbert space: either $ \mc{H}_{\al}^{(\unl{0})} $ or $ \mc{H}^{(\underline{m})}_{\al } $ which are subspaces of the Hilbert space $  \mc{H}_{\al}:=\mc{H}_{\al}^{(\unl{0})}\otimes_{(\underline{m})}\mc{H}^{(\underline{m})}_{\al } $ . 

For each cell, there is a basis of eigenstates that diagonalizes the corresponding scalar field operators: $ \ \hat{\!\phi}^{(\unl{0})}_{\al}(t)\equiv\hat{\!\phi}^{(\unl{0})}(\boldsymbol{x}_{\al},t)$ and $\ \hat{\!\phi}^{(\unl{m})}_{\al }(t)\equiv\ \hat{\!\phi}^{(\unl{m})}(\boldsymbol{x}_{\al},t) $, namely
\begin{equation}\label{BasisofHm}
 \ \hat{\!\phi}^{(\boldsymbol{m})}_{\al}(t)\ket{\phi^{(\boldsymbol{m})}_{\al},t}   :={\phi}^{(\boldsymbol{m})}_{\al }\, \ket{\phi^{(\boldsymbol{m})}_{\al},t}\ ,
\end{equation}
where we have introduced the bold symbol $ {\small(\boldsymbol{m})} $ to succinctly  mean $ (\unl{0}) $ and $  (\unl{m}) $. The product
\begin{align*}
\ket{\phi^{(\unl{0})}_{\al},t}\ket{\phi^{(1,0,\ldots,0)}_{\al},t}\ket{\phi^{(2,0,\ldots,0)}_{\al},t} \cdots\ket{\phi^{(m_{1},m_{2},\ldots,m_{n})}_{\al},t}\cdots\\
=\ket{\phi^{(\unl{0})}_{\al},t}\prod_{(\unl{m})}\ket{\phi^{(\unl{m})}_{\al},t}\equiv\ket{\prod_{(\boldsymbol{m})}\phi^{(\boldsymbol{m})}_{\al},t}=: \ket{\Phi_{\al },t}
\end{align*}
 is a basis for $ \mc{H}_{\al } $ that in fact diagonalizes the linear operator  $\wht{\Phi}_{\al \bt}$. Indeed, a direct calculation leads to
\begin{equation}\label{BasisforH}
\wht{\Phi}_{\al \bt}(t)\ket{\Phi_{\al },t}={\Phi}_{\al \bt}\ket{\Phi_{\al },t}\ .
\end{equation}
Notice that the $ \bt- $index content of the operator on the left hand side comes from the argument in the cosines, see\reff{FiopFourier}, therefore on the right hand side of Eq.~\eqref{BasisforH}, such content is present only in the corresponding  eigenvalues. The  orthonormality and completeness relations for the basis $ \ket{\Phi_{\al},t} $,
\begin{subequations}\label{ortho-compl-Phi}
	\begin{align}
	\brkt{\Phi_{\al},t}{\Phi_{\al'},t}	& = \dlt_{\al\al'} \ ,\label{orthoPhi}\\
	\int\ud \Phi_{\al}	\ket{\Phi_{\al},t}\bra{\Phi_{\al},t}& = 1 \ ,\label{}
	\end{align}
\end{subequations}
are consistent with the corresponding relations
\begin{subequations}\label{complete-ortho}
	\begin{align}
	\brkt{\phi^{(\boldsymbol{m})}_{\al},t}{\phi^{(\boldsymbol{n})}_{\al'},t} & = \dlt_{\al\al'}^{(\boldsymbol{mn})}, \  \label{ortho3}\\
 	\int \ud\phi^{(\boldsymbol{m})}_{\al}\ket{\phi^{(\boldsymbol{m})}_{\al},t}\bra{\phi^{(\boldsymbol{m})}_{\al},t} & = 1\ ,\ \trm{(no sum)}\label{compl2}
 \end{align}
\end{subequations}
respectively, provided that
\begin{equation}\label{measures}
\int\ud \Phi_{\al}=\int\ud\phi^{(\unl{0})}_{\al}\prod_{(\unl{m})}\int\ud\phi^{(\unl{m})}_{\al} \equiv \prod_{(\boldsymbol{m})} \int\ud\phi^{(\boldsymbol{m})}_{\al}.
\end{equation}
In Eq.~\eqref{ortho3} the multindex Kronecker delta $ \dlt^{(\boldsymbol{mn})}_{\al\al'} $ numerically equals to one whenever $ \al=\al' $ and the arrangement of numbers $ (\boldsymbol{m}) $ coincides with $ (\boldsymbol{n}) $, and vanishes otherwise.

The transition amplitude\reff{FKernel} can be expressed in terms of the amplitudes at each cell. Using the orthonormality relations Eqs.\eqref{ortho3},
\begin{equation}\label{FKernel2}
\brkt{\Phi',t'}{\Phi,t}  \simeq\!\prod_{\al=1}^{M}\brkt{\Phi^{\prime}_{\al},t'}{\Phi_{\al},t}=\!\prod_{\al=1}^{M}\prod_{(\boldsymbol{m})}\brkt{\phi^{(\boldsymbol{m})\prime}_{\al},t'}{\phi^{(\boldsymbol{m})}_{\al},t}
\end{equation}
where the approximation symbol can be replaced by an equality only after taking the limit $ M\to \infty $ under the condition $ M\Dlt V=  V $ fixed, and using $ V\to\infty $ \emph{a posteriori}.  It is convenient now to consider the split of the time interval  into $ N $ segments $ t_{l} $, $ l=0,1,\ldots,N$, so that $ t_{0}=t, t_{1}=t+\eps, \ldots,  t_{N}=t' $ and to use the completeness relations Eqs.~\eqref{compl2} at the time slices, so that, 
\begin{align}\label{FKernel3}
& \brkt{\Phi',t'}{\Phi,t}\simeq\prod_{(\boldsymbol{m})}\int\ud\phi^{(\boldsymbol{m})}_{\al {N-1}}\cdots \ud\phi^{(\boldsymbol{m})}_{\al {1}}  \nonumber\\
& \times \brkt{\phi^{(\boldsymbol{m})\prime}_{\al},t'}{\phi^{(\boldsymbol{m})}_{\al N-1},t_{N-1}}
\cdots
\brkt{\phi^{(\boldsymbol{m})}_{\al 1},t_{1}}{\phi^{(\boldsymbol{m})}_{\al},t}\ .
\end{align}
The infinitesimal transition amplitude for each mode is of the form
\begin{align}\label{infKernel}
\brkt{\phi^{(\boldsymbol{m})}_{\al l+1},t_{l+1}}{\phi^{(\boldsymbol{m})}_{\al l},t_{l}}& =\bra{\phi^{(\boldsymbol{m})}_{\al l+1}}e^{-i\eps\wht{H}}\ket{\phi^{(\boldsymbol{m})}_{\al l}}\nonumber\\
& \simeq\bra{\phi^{(\boldsymbol{m})}_{\al l+1}}1-i\eps\wht{H}\ket{\phi^{(\boldsymbol{m})}_{\al l}}+\mc{O}(\eps^{2})\ .
\end{align}
where $ \wht{H} $ is the Hamiltonian operator that provides the time evolution associated to the theory \eqref{Xdim-action} on the spacetime grid. In terms of the operator-valued Fourier modes,  it can be written as follows:
\begin{equation}\label{Hdiscrete}
\wht{H}=\sum_{\alpha=1}^{M}\Dlt V \,\wht{\mc{H}}\left(\hat{\pi}^{(\unl{0})}_{\al}, \hat{\pi}^{(\unl{m})}_{\al},\ \hat{\!\phi}^{(\unl{0})}_{\al},\hat{\!\phi}^{(\unl{m})}_{\al}\right)\ .
\end{equation}
At this point the conventional way to calculate a matrix element \eqref{infKernel} applies (see for example\citte{grebook}). So,  conveniently introducing a complete set of momentum eigenstates (for each mode) at each time slice, 
\begin{equation}
		\int\frac{\Dlt V\ud\pi^{(\boldsymbol{m})}_{\al l}}{2\pi}\ket{\pi^{(\boldsymbol{m})}_{\al l},t_{l}}\bra{\pi^{(\boldsymbol{m})}_{\al l},t_{l}}  = 1\ ,\label{picompl2}
\end{equation}
and using the notation $ \mc{H}_{\al l}:=\mc{H}\left({\pi}^{(\unl{0})}_{\al l}, {\pi}^{(\unl{m})}_{\al l}, {\phi}^{(\unl{0})}_{\al l},{\phi}^{(\unl{m})}_{\al l}\right) $ for the Hamiltonian density on the grid, one obtains
	\begin{align}\label{PSdiscreteK}
	\brkt{\Phi',t'}{\Phi,t}	& \simeq \!\prod_{(\boldsymbol{m})} \prod_{\alpha=1}^{M}\left(\prod_{l=1}^{N-1}\int \ud\phi^{(\boldsymbol{m})}_{\al l}\cdot\prod_{k=0}^{N-1}\int \frac{\Dlt V\ud\pi^{(\boldsymbol{m})}_{\al k}}{2\pi}\right)\nonumber\\
		& \times\exp\left[i\sum_{l=0}^{N-1}\eps\sum_{\al=1}^{M}\Dlt V \left(\pi^{(\unl{0})}_{\al l}\frac{\phi^{(\unl{0})}_{\al l+1}-\phi^{(\unl{0})}_{\al l}}{\eps}+\right.\right.\nonumber\\
		& \left. \left. \sum_{(\unl{m})}\pi^{(\unl{m})}_{\al l}\frac{\phi^{(\unl{m})}_{\al l+1}-\phi^{(\unl{m})}_{\al l}}{\eps}-\mc{H}_{\al l}\right)\right] .
	\end{align}
In the limits $ N\to \infty $, $ M\to \infty $ and $ V\to \infty $,  this object defines the Feynman kernel  as a phase space path integral
	\begin{align}\label{PSKernel}
&	\brkt{\Phi',t'}{\Phi,t}	 = \prod_{(\boldsymbol{m})}\int\mc{D}\phi^{(\boldsymbol{m})}\mc{D}\pi^{(\boldsymbol{m})}\nonumber\\	
	& \times\exp\left[i\int_{t}^{t'}\ud t\int\ud^{3}x\left(\pi^{(\unl{0})}\dot{\phi}^{(\unl{0})}+\sum_{(\unl{m})}\pi^{(\unl{m})}\dot{\phi}^{(\unl{m})}\right.\right.\nonumber\\
	& \left.\ph{\int_{t}^{t'}}\left.\ph{\sum_{\unl{m}}}-\mc{H}(\pi^{(\unl{0})},\phi^{(\unl{0})},\pi^{(\unl{n})},\phi^{(\unl{n})})\right)\right] . 
\end{align}

A typical Lagrangian density,
\begin{equation*}
\mc{L}_{(4+n)}=\dfrac{1}{2}\left(\partial_{M}\Phi\cdot\partial^{M}\Phi-m^{2}\Phi^{2}\right)-V_{(4+n)}(\Phi)\,,
\end{equation*} 
leads to a Hamiltonian density, $ \mc{H}(\pi^{(\unl{0})},\phi^{(\unl{0})},\pi^{(\unl{m})},\phi^{(\unl{m})}) $, quadratic in momenta, so that, after performing Gaussian integrals, Eq. \eqref{PSKernel} can be written as
\begin{equation}\label{CSFeynmanKernel}
	\brkt{\Phi',t'}{\Phi,t}	 = \mc{K}\prod_{(\boldsymbol{m})} \int\mc{D}\phi^{(\boldsymbol{m})}e^{i\int_{t}^{t'}\!\!\ud t L}\ ,	
\end{equation}
here $\mc{K} $ is a field independent constant and $ L=\int \ud^{3}x\mc{L} $. Notice that $ \mc{L} $ is a Lagrangian density whose fields are defined on $ \mc{M}^{4} $, since it corresponds to $ \int_{\mc{N}^{n}}\mc{L}_{(4+n)} $. The Lagrangian $ \mc{L} $ is the  dimensionally  reduced description on $\mc{M}^{4}$ from the higher dimensional correlative theory.

We have shown that $\brkt{\Phi',t'}{\Phi,t}$ can be expressed either as $\int\mc{D}\Phi \exp[i\int\ud t\,\mc{L}_{(4+n)}(\Phi,\partial\Phi)]$ or Eq. \eqref{CSFeynmanKernel}, hence we conclude that the functional measure is invariant under the Kaluza Klein compactification process. According to\citte{unz85} this means that KK dimensional reduction is quantum mechanically consistent, even after formally considering \emph{all} KK modes.

\section{The $k$-point Green's functions: Generalities}
\label{sec:Green}

As it is well known, all $ k $-point Green's functions, connected graphs and physical quantities can be obtained from the vacuum-to-vacuum transition in the presence of sources ${Z}[J]\propto \brkt{0,\infty}{0,-\infty}^{J} $. Sources are coupled to fields only, but momenta, hence for a typical Lagrangian this amplitude is proportional to\reff{CSFeynmanKernel} after adding $ \int\ud^{3}x[J^{(\unl{0})}\phi^{(\unl{0})}+\sum_{(\unl{m})}J^{(\unl{m})}\phi^{(\unl{m})}]\equiv\sum_{(\boldsymbol{m})}J^{(\boldsymbol{m})}\phi^{(\boldsymbol{m})} $ to the Lagrangian density and taking the limits $ t\to-\infty $ and $ t'\to\infty $. Therefore,
\begin{equation}\label{WJ}
 Z[J^{(\boldsymbol{m})}]= \mc{K}\prod_{(\boldsymbol{m})} \int\mc{D}\phi^{(\boldsymbol{m})}\ e^{\ i\int\!\ud^{4} x \left[\mc{L}+\sum_{(\boldsymbol{m})}J^{(\boldsymbol{m})}\phi^{(\boldsymbol{m})}\right]}\ .	
\end{equation}

In this amplitude the zero and the excited modes decouple from each other for the free Lagrangian case, \ie $ \mc{L}_{0}:=\mc{L}|_{V(\phi^{(\unl{0})},\phi^{(\unl{m})})=0} $. Denoting this case by $ Z_{0} $, one has
\begin{equation}\label{Wo}
	Z_0\left[J^{(\boldsymbol{m})}\right]=\prod_{(\boldsymbol{m})}Z_0^{(\boldsymbol{m})}\left[J^{(\boldsymbol{m})}\right],
\end{equation}      
which, after neglecting needless multiplicative factors, is nothing but the multiplication of the free generating functionals
\begin{equation}\label{Wo-n}
Z_0^{(\boldsymbol{m})}\!\left[J^{(\boldsymbol{m})}\right]\!\! =\exp\left[-\frac{i}{2}\int d^4y d^4x J^{(\boldsymbol{m})}(y)\Delta^{(\boldsymbol{m})}_{yx}J^{(\boldsymbol{m})}(x)\right].
\end{equation}
The Feynman propagators involved here are 
\begin{equation}\label{FeynProps}
 \Delta^{(\boldsymbol{m})}_{xy} = \int\frac{\ud^{4} k}{(2\pi)^{4}} \frac{e^{ik(x-y)}}{k^{2}-m^{2}_{{(\boldsymbol{m})}}+i\varepsilon}\ 
\end{equation}
where $m^{2}_{{(\unl{0})}} = m^{2}$ and $ m^{2}_{{(\unl{m})}} = m^{2}+\left(\frac{\unl{m}_{1}}{R_{1}}\right)^{2}+\cdots+\left(\frac{\unl{m}_{n}}{R_{n}}\right)^{2}$. The naked masses $ m^{2}_{{(\unl{m})}} $ are different from the squared mass parameter $ m^{2} $ in all cases, for {\tiny }example, if the number of extra dimensions is $ n=3 $ and the arrangement $ (\unl{m})=(m_{1},0,m_{3}) $, for fixed integers $ m_{i} $ (which are necessarily nonzero), we have $ m^{2}_{{(\unl{m})}}= m^{2}+\left(\frac{{m}_{1}}{R_{1}}\right)^{2}+\left(\frac{{m}_{3}}{R_{3}}\right)^{2}\ $.

Reconsidering\reff{WJ} we formally write it as follows:
\begin{align}\label{WJ2}
	 & Z[J^{(\boldsymbol{m})}]\propto  \exp\left[-i\int\ud^{4}x\,V\left(\frac{1}{i}\frac{\dlt}{\dlt J^{(\boldsymbol{m})}(x)}\right)\right] \nonumber\\ 
	&\times \prod_{(\boldsymbol{m})} \int\mc{D}\phi^{(\boldsymbol{m})}\exp\left[i\int\!\ud^{4} x \left(\mc{L}_{0}+\sum_{(\boldsymbol{m})}J^{(\boldsymbol{m})}\phi^{(\boldsymbol{m})}\right)\right]\	,
\end{align}
for well behaved potential functions, or equivalently using Eqs.~\eqref{Wo} and \eqref{Wo-n}
\begin{equation}\label{WJ3}
Z[J^{(\boldsymbol{m})}]\propto  \exp\left[-i\int\ud^{4}x\,V\left(\frac{1}{i}\frac{\dlt}{\dlt J^{(\boldsymbol{m})}(x)}\right)\right]Z_{0}[J^{(\boldsymbol{m})}]\ .
\end{equation}
The normalization factor is chosen as the right hand side of the expression\reff{WJ3} evaluated at $ J^{(\boldsymbol{m})}\!\equiv 0 $, such factor has the effect of removing vacuum bubble graphs. As it is well known, $ Z[J^{(\boldsymbol{m})}] $ is the generator of the different Green's functions of the theory when written as a Volterra series, that is,
\begin{align}\label{volterra}
Z & =  \sum_{k=0}^{\infty}\int\!\ud x_{1}\cdots\ud x_{k}\frac{i^{k}}{k!}\!\sum_{({\boldsymbol{m}_{1}}\cdots\, \boldsymbol{m}_{k})}\!\!\! G^{(\boldsymbol{m}_{1}\cdots \boldsymbol{m}_{k})}(x_{1},\ldots, x_{k})\nonumber\\
	&\ \ \times J^{(\boldsymbol{m}_{1})}(x_{1})\cdots\,J^{(\boldsymbol{m}_{k})}(x_{k})
\end{align}
where 
\begin{equation}\label{GFs}
G^{(\boldsymbol{m}_{1}\cdots \boldsymbol{m}_{k})}(x_{1},\ldots, x_{k})\! :=\frac{\dlt^{k}Z}{i^{k}\dlt J^{(\boldsymbol{m}_{1})}(x_{1})\cdots\,\dlt J^{(\boldsymbol{m}_{k})}(x_{k})}\Bigg\rvert_{J=0}
\end{equation}	
are the $ k $-point Green's functions whose specific form hinges on the structure of the different interactions in the Lagrangian. 

\begin{figure}[h]
\includegraphics[width=6.5cm, height=5.5cm]{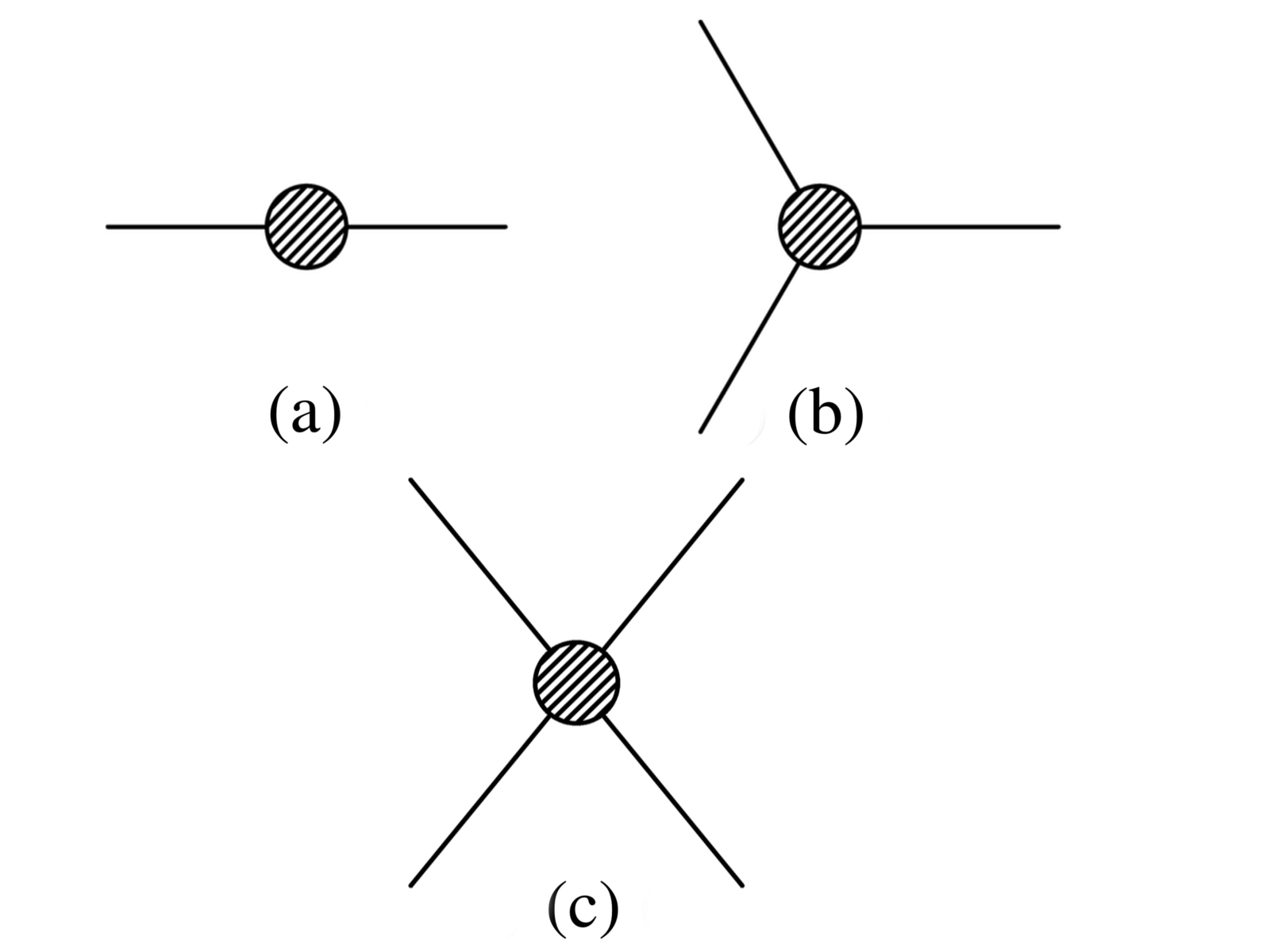}
\caption{Examples of Standard Green's Functions: \hspace{2cm} \textbf{(a)} $G^{(\unl{0}\unl{0})}$, \textbf{(b)} $ G^{(\unl{0}\unl{0}\unl{0})} $, \textbf{(c)} $ G^{(\unl{0}\unl{0}\unl{0}\unl{0})} $; solid line stands for the zero mode $(\unl{0})$ propagation.}
\label{fig:SGF}
\end{figure}

\begin{figure}[h]
\includegraphics[width=6cm, height=5cm]{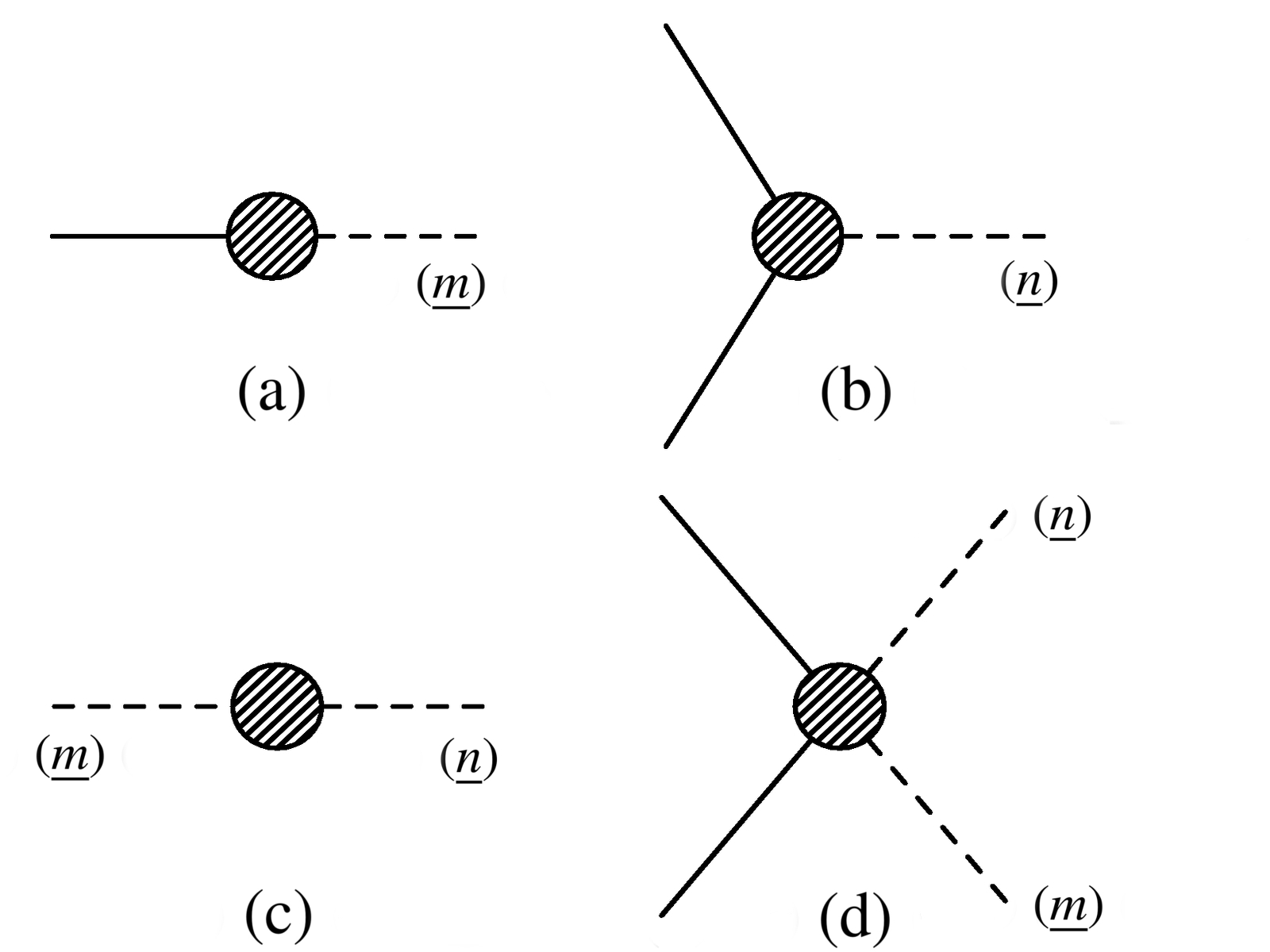}
\caption{Examples of Non-Standard Green's Functions: \hspace{2cm} \textbf{(a)} $G^{(\unl{0}\unl{m})}$, \textbf{(b)} $ G^{(\unl{0}\unl{0}\unl{n})} $, \textbf{(c)} $ G^{(\unl{m}\unl{n})} $, \textbf{(d)} $ G^{(\unl{0}\unl{0}\unl{m}\unl{n})} $; dashed line stands for excited model $(\unl{m})$ propagation.}
\label{fig:NSGF}
\end{figure}

In the expressions \eqref{volterra} and \eqref{GFs} each index $ (\boldsymbol{m}_{i}) $ is a condensed label that takes the different values $ (\unl{0}) $ and $ (\unl{m}_{i}) $, then $G^{(\boldsymbol{m}_{1}\cdots \boldsymbol{m}_{k})}$ symbolizes essentially $ (k+1) $ different types of Green's functions. For example, there are essentially three different types of 2-point functions $G^{(\boldsymbol{mn})}$: $ G^{(\unl{0}\unl{0})} $, $ G^{(\unl{0}\unl{n})}$ and  $ G^{(\unl{m}\unl{n})}$; essentially four different 3-point functions $G^{(\boldsymbol{lmn})}$: $ G^{(\unl{0}\unl{0}\unl{0})} $, $ G^{(\unl{0}\unl{0}\unl{n})}$, $ G^{(\unl{0}\unl{m}\unl{n})}$, $ G^{(\unl{l}\unl{m}\unl{n})}$; and so on. The $G^{(\unl{0}\unl{0})},\ldots,  G^{(\unl{0}\cdots\unl{0})} $ are called Standard Green's Functions (SGFs), whose external legs exclusively correspond to the zeroth mode field, see \textbf{Figure ~\ref{fig:SGF}},  whereas the rest essentially different are called Non-Standard Green's Functions (NSGFs)\citte{cor13}, see \textbf{Figure ~\ref{fig:NSGF}}. In the case where no extra dimensions are present, $(\unl{0})$ collapses to $(0)$ and $(\unl{m})$ does not exist at all, hence only the usual GFs do exist. In general, depending on the internal structure of the Lagrangian, some of the NSGFs  may vanish.

In the noninteracting theory, see\reff{Wo}, the corresponding Green's functions can be calculated explicitly. The $ (2k+1) $-point Green's functions vanish. The $(2k)$-point Green's functions correspond to the free propagation of massive scalar free particles: $ G^{(\unl{00})}_{0} $ and $ G^{(\unl{mm})}_{0} $ are interpreted as the free propagation of one `zeroth mode' particle, \textbf{Figure ~\ref{fig:SGF}a}, and one `excited mode' particle, \textbf{Figure ~\ref{fig:SGF}b}, respectively. The functions $ G^{(\unl{0n})}_{0} $ vanish conveying the information that there is no possibility of interchanging the nature of the particle during a free propagation; the non-vanishing  $ G^{(\unl{0000})}_{0} $, $ G^{(\unl{00nn})}_{0} $, and $ G^{(\unl{mmnn})}_{0} $ are interpreted as the free propagation of two free massive particles, that of  two `zeroth mode' particles, one `zeroth mode' and one `excited mode' particles (\textbf{Figure ~\ref{fig:SGF}c}), and two `excited mode' particles, respectively; and so on. From\reff{FeynProps} it can be read that any excited KK field is heavier than the zeroth mode, and the smaller the extra dimensions the heavier they become; for this reason, given an integer $n$, the fields are separated into the light field $\phi^{(\unl{0})}$ and the heavy fields $\phi^{(\unl{m})}$.

\begin{figure}
\includegraphics[width=6cm, height=4.5cm]{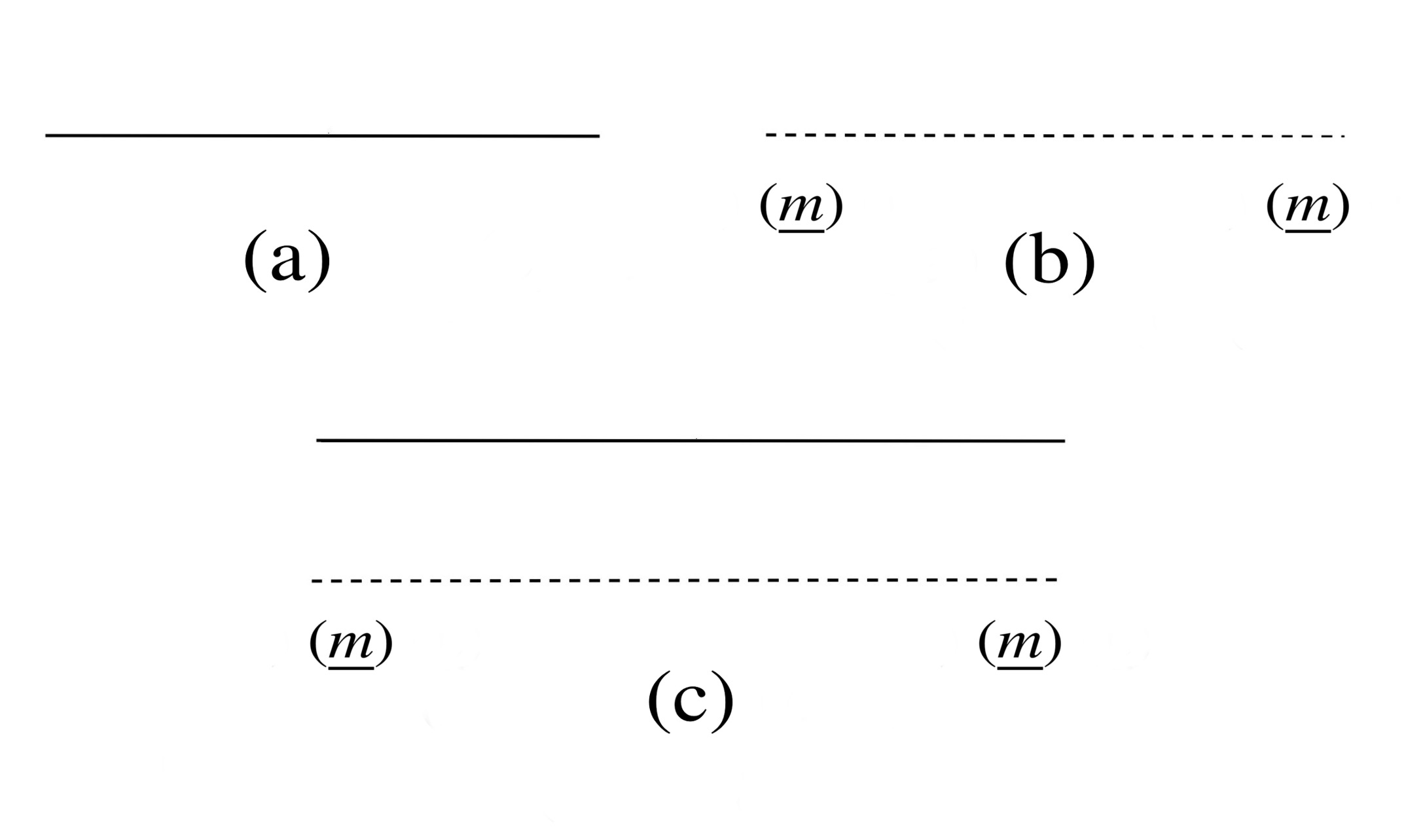}
\caption{Free Propagators}
\label{fig:Free}
\end{figure}

The interacting case, see\reff{WJ3}, is endowed with perturbation character when the operator preceding $ Z_{0}$ is expanded with respect to very weak coupling constants. It is worth remarking that even if one keeps such expansion to a finite order, each $ k $-point Green's function still contains an infinite number of diagrams due to the presence of the infinite number of interactions in the Lagrangian $ \mc{L} $, either between zeroth and excited Kaluza Klein (KK) modes, or, among KK modes only [\eg\reff{Lcfir}]. This fact brings issues in regularization procedures. 

The generating functional of connected Green's functions is
\begin{equation}\label{ZJoJm}
W[J^{(\boldsymbol{m})}]:=-i\ln \lbrace Z[J^{(\boldsymbol{m})}]\rbrace
\end{equation}
where
\begin{equation}\label{Gc}
G^{(\boldsymbol{m}_{1}\cdots \boldsymbol{m}_{k})}_{c}(x_{1},\ldots, x_{k})\! :=\frac{\dlt^{k}W}{i^{k}\dlt J^{(\boldsymbol{m}_{1})}(x_{1})\cdots\,\dlt J^{(\boldsymbol{m}_{k})}(x_{k})}\Bigg\rvert_{J=0}
\end{equation}
are the connected Green's functions of the theory.

\section{Self-interacting field in extra dimensions and its effective description}
\label{secc:model}

In this section we consider the case $ V_{(4+n)}(\Phi)=\frac{\lm_{(4+n)}}{r!}\Phi^{r} $  and the corresponding dimensionally reduced La\-gran\-gian $ \mc{L} $. The effective four dimensional Lagrangian can be written as the sum of three qualitatively different sectors,
\begin{equation}\label{Lcfir}
\mc{L}=\mc{L}^{(\unl{0})}+\sum_{(\unl{m})}\mc{L}^{(\unl{0m})}+\sum_{(\unl{m})}\mc{L}^{(\unl{m})}\, .
\end{equation}
Namely, the purely light field $  \phi^{(\unl{0})}\equiv\phi $ self-interacting model,
\begin{subequations}\label{Lfir}
\begin{equation}\label{L0fir}
\mc{L}^{(\unl{0})}=\frac{1}{2}\left((\partial_{\mu}\phi)(\partial^{\mu}\phi)-m^{2}\phi^{2}\right)-\frac{\lm}{r!}\phi^{r}\, ,
\end{equation}
the interaction terms between light and heavy fields,
\begin{align}\label{L0mfir}
\mc{L}^{(\unl{0m})}=& -\frac{\lm}{r!}\sum_{K=2}^{r-1}  {{r}\choose{K}}\phi^{r-K}\phi^{(\unl{m})}\nonumber\\
&\times\sum_{(\unl{m_{1}})}\cdots\sum_{(\unl{m_{K-1}})}\phi^{(\unl{m_{1}})}\cdots\phi^{(\unl{m_{K-1}})}\Dlt_{(\unl{mm_{1}\cdots m_{K-1}})},
\end{align}
and the purely excited KK sector
\begin{align}\label{Lmfir}
\mc{L}^{(\unl{m})}= & \frac{1}{2}\left(\partial_{\mu}\phi^{(\unl{m})}\partial^{\mu}\phi^{(\unl{m})}-m^{2}_{{(\unl{m})}}\phi^{(\unl{m})}\phi^{(\unl{m})}\right)-\frac{\lm}{r!}\phi^{(\unl{m})}\nonumber\\
& \times\sum_{(\unl{m_{1}})}\cdots\sum_{(\unl{m_{r-1}})}\phi^{(\unl{m_{1}})}\cdots\phi^{(\unl{m_{r-1}})}\Dlt_{(\unl{mm_{1}\cdots m_{r-1}})} .
\end{align}
\end{subequations}
In the set of Eqs.~\eqref{Lfir}, the dimensionless multi-index  $ \Dlt_{(\cdots)} $ is defined as follows:
\begin{equation}\label{Lmdfir}
\Dlt_{(\unl{m}\cdots\unl{l})}:=\frac{2^{K/2}}{\prod_{i=1}^{n}2\pi R_{i}}\int\ud^{n}\bar{x}\cos\left(\frac{\unl{m}\bar{x}}{R}\right)\cdots\cos\left(\frac{\unl{l}\bar{x}}{R}\right)\ ,
\end{equation}
for $K$ indices, and the universal coupling constant $	\lm  $ is defined as $ \lm_{(4+n)}/\left( \prod_{i}2\pi R_{i} \right)^{\frac{r}{2}-1}$.  In the Dyson's sense, cases $ r>4 $ are non-renormalizable, $ r=3 $ becomes super-renormalizable, and the case $ r=4 $ is renormalizable. This renormalizability, as we know, makes reference to divergences that arise due to short-distance effects; nevertheless, the presence of an infinite number of interactions between the zeroth mode and KK modes induces another type of divergence. 

Of special interest for us is the case $ r=4 $. The corresponding sectors in\reff{Lcfir} consist of $ (i) $ the well-known $ \phi^{4} $ self-interacting model,
\begin{subequations} 
\begin{equation}\label{L0}
\mc{L}^{(\unl{0})}=\frac{1}{2}\left((\partial_{\mu}\phi)(\partial^{\mu}\phi)-m^{2}\phi^{2}\right)-\frac{\lm}{4!}\phi^{4}\, ,
\end{equation}
$ (ii) $ the sector that couples the light and heavy fields,  
\begin{equation}\label{L0m}
\mc{L}^{(\unl{0m})}=-\frac{\lm}{12}\phi\left(3\phi \phi^{(\unl{m})2}+2\phi^{(\unl{m})}\sum_{(\unl{ln})}\phi^{(\unl{l})}\phi^{(\unl{n})}\Dlt_{(\unl{mln})}\right) ,
\end{equation}
and $ (iii) $ the pure KK excited sector,
\begin{align}\label{Lm}
\mc{L}^{(\unl{m})}= & \frac{1}{2}\left(\partial_{\mu}\phi^{(\unl{m})}\partial^{\mu}\phi^{(\unl{m})}-m^{2}_{{(\unl{m})}}\phi^{(\unl{m})}\phi^{(\unl{m})}\right)\nonumber\\
& -\frac{\lm}{4!}\phi^{(\unl{m})}\sum_{(\unl{lnp})}\Dlt_{(\unl{mlnp})}\phi^{(\unl{l})}\phi^{(\unl{n})}
\phi^{(\unl{p})}\, .
\end{align}
\end{subequations}

\begin{figure}[h]
\includegraphics[width=6cm, height=2cm]{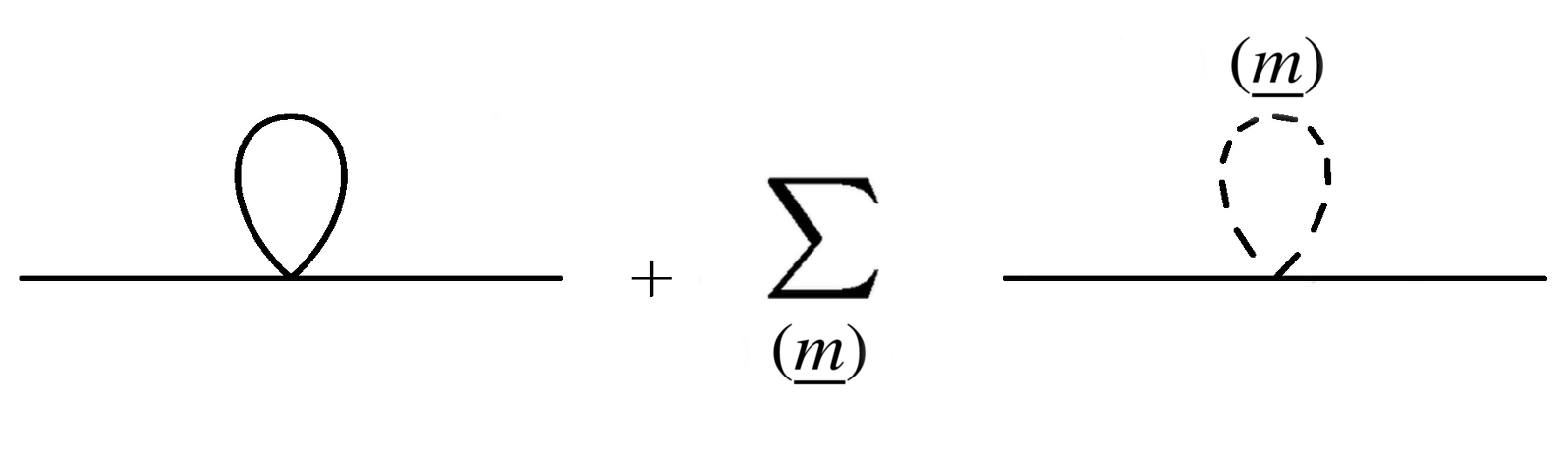}
\caption{Contributions to the two point SVF $\Gamma_2(p)$.}
\label{fig:mass-self}
\end{figure}

In our construction, the extra-dimensional effects remain encoded in the excited KK modes. Due to  the presence of the terms $ \sum_{(\unl{m})}\mc{L}^{(\unl{0m})} $, there are extra dimensional contributions to the SGFs. In turn they modify standard vertex functions (SVFs); for example, to the 1-loop approximation, the two point SVF $\Gamma_{2}(p)$ becomes
\begin{equation}
\Gamma_{2}(p)=p^2-m^2-\left(M^{(\unl{0})}(p^2)+\sum_{(\underline{m})}M^{(\underline{m})}(p^2)\right),\label{2PVF}
\end{equation}
where $ \Gamma_{2}(p) G^{(\unl{0}\unl{0})}_{c}(p)=i$. There are two contributions to the self-energy, one from the zeroth mode $M^{(\unl{0})}(p^2)$ and  the other from the sum of excited modes $M^{(\underline{m})}(p^2)$, see \textbf{Figure ~\ref{fig:mass-self}}; they come from the $\phi^{4}$ self-interaction term in $ \mc{L}^{(\unl{0})} $ and the first interaction term in $ \mc{L}^{(\unl{0m})} $,\reff{L0m}, respectively. The corresponding expressions are condensed in the following expression
%\begin{subequations}\label{MSE}
%\begin{equation}\label{M0SE}
%-iM^{(\underline{0})}=-i\frac{\lambda}{2}\int{\frac{d^4k}{(2\pi)^4}}\frac{i}{k^2-m_{\phi^{(\underline{0})}}^2},
%\end{equation}
\begin{equation}\label{MmSE}
M^{(\boldsymbol{m})}=\frac{\lambda}{2}\int{\frac{d^4k}{(2\pi)^4}}\frac{i}{k^2-m_{{(\boldsymbol{m})}}^2}\ .
\end{equation}
%\end{subequations}
From an effective point of view, these two contributions impact the physical mass of the light field. 

\begin{figure}[h]
\includegraphics[width=5cm, height=2cm]{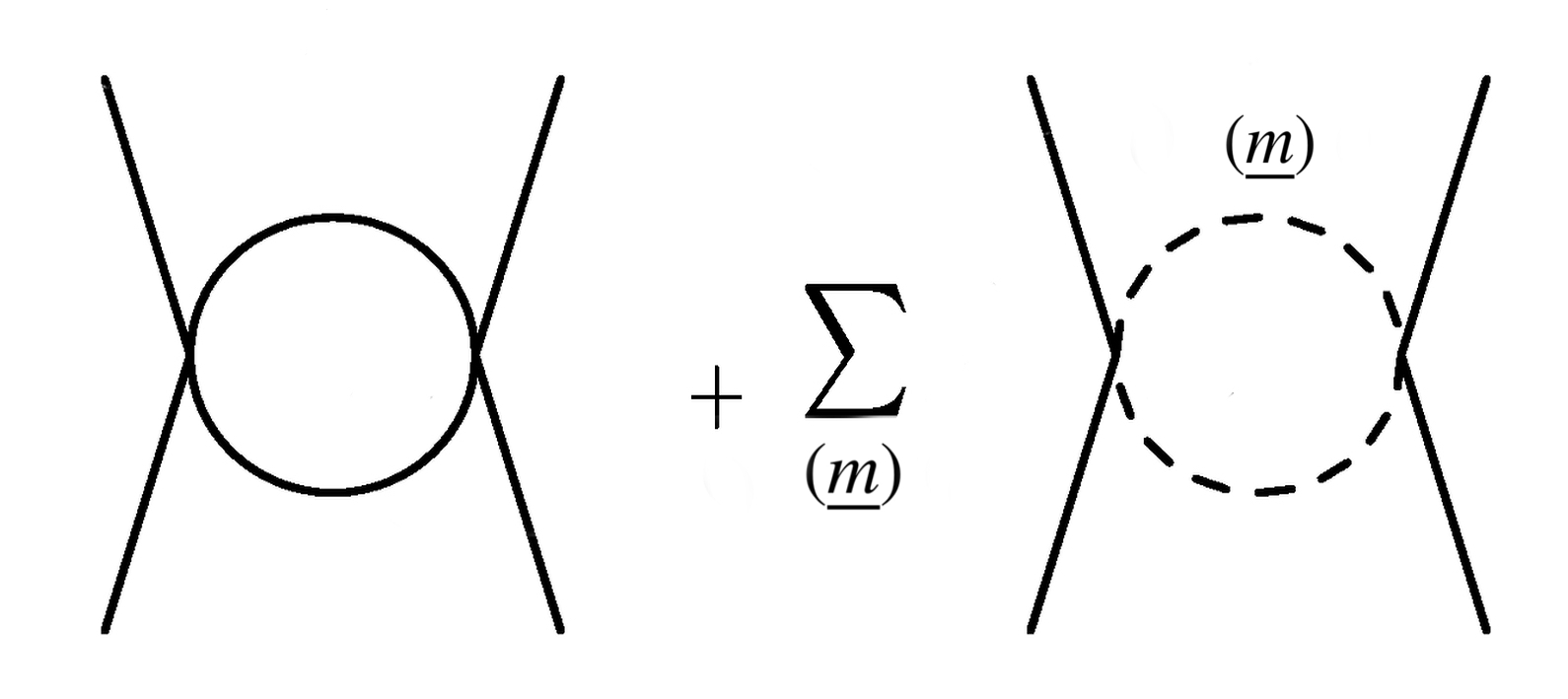}
\caption{Contributions to the four point SVF $\Gamma_4(p_i)$.}
\label{fig:four-vertex}
\end{figure}

In a similar fashion, the four point SVF defined by 
$ \Gm_{4}(p_{i})=G^{(\unl{00})}(p_{1})^{-1} G^{(\unl{00})}(p_{2})^{-1}$ $G^{(\unl{00})}(p_{3})^{-1}$ $G^{(\unl{00})}(p_{4})^{-1}$ $\times  G^{(\unl{0000})}(p_{i}) $
to the  1-loop approximation, becomes
\begin{equation}\label{4PVF}
\Gamma_{4}(p_{i})=-i\lambda+\left(\Gamma_{\trm{1-loop}}^{(\underline{0})}(p_{i})+\sum_{(\underline{m})}\Gamma_{\trm{1-loop}}^{(\underline{m})}(p_{i})\right)
\end{equation}
where $ \Gamma_{\trm{1-loop}}^{(\underline{0})} $ and $\Gamma_{\trm{1-loop}}^{(\underline{m})}$ are correlated to the zeroth and a KK excited mode particle circulating within the loop, respectively, see \textbf{Figure ~\ref{fig:four-vertex}}. Summing over the Maldestam variables each of these contributions can be summarized as follows:
%\begin{subequations}\label{Gv4}
	\begin{equation}
\Gamma_{\trm{1-loop}}^{(\boldsymbol{m})}(p_{i})=\sum_{\lbrace p\rbrace} \Delta\Gamma^{(\boldsymbol{m})}(p)\label{Gv4pm}
	\end{equation}
%\end{subequations}
where the symbol $\sum_{\lbrace p\rbrace}$ indicates a sum over the three Maldestam variables, and 
%\begin{subequations}\label{Deltas}
	\begin{equation}
\Delta\Gamma^{(\boldsymbol{m})}(p) =-\frac{1}{2}\lambda^2\int \frac{d^4k}{(2\pi)^4}\frac{i}{k^2-m_{{(\boldsymbol{m})}}^2}\frac{i}{(p-k)^2-m_{{(\boldsymbol{m})}}^2}.\label{DeltaGm}
	\end{equation}
%\end{subequations}

It is now explicit that\reff{2PVF} and\reff{4PVF} will present divergences due to the internal momentum integrals \emph{and} the infinite sum over them.

\section{Final remarks}
\label{secc:fr}

In this paper the dimensional reduction of a scalar field theory has been done at the level of the Feynman path integral. We showed that, even if one keeps the whole KK towers, the functional measure is invariant under the Kaluza-Klein compactification process. Which me\-ans that KK dimensional reduction is quantum mechanically consistent, even after formally considering \emph{all} KK modes. 

We analyzed the generalities of Green's functions of a general effective scalar theory.  The free generating functional $Z_{0}$ for the effective scalar free theory  can be written as a Volterra series where Green's functions are interpreted as the propagation of  free either light or heavy particles. In a self-interacting scalar case one can see that each $k$-point Green's function contains an infinite number of diagrams, even at a finite-loop aproximation, the source being the infinite number of interactions in the Lagrangian density. So, even assuming that the dimensionally reduced theory is renormalizable, in the Dyson's sense, it is not immediate that the infinite number of contributions  to each order in perturbation theory can be treated. 

Towards a plausible regularization process, we notice that $\Gm_{2}$ and $\Gm_{4}$, Eqs.~\eqref{2PVF} and \eqref{4PVF}, have the following form after assuming $R_{1}=\cdots =R_{n}\equiv R$  and applying dimensional regularisation techiniques:
\begin{align}
\Gamma_{2}=&\ p^2-m^2-\frac{\lambda}{32\pi^{2}}m^{2}\left(\frac{4\pi\mu^{2}}{m^{2}}\right)^{1+s}\Bigg[\Gamma\left(s\right)\notag\\
& \left.+\sum_{l=1}^{n}\binom{n}{l}c^{2s}E_l^{c^2}(s)\Gamma(s)\right]\label{2PVFReg}
\end{align}
and
\begin{align}\label{4SVFDR}
\Gamma_{4} & = -i\lambda\mu^{2s}+\frac{i\lm^{2}\mu^{2s}}{32\pi^{2}} \sum_{\lbrace p \rbrace} \int_{0}^{1}\ud z\left[\Gm(s)\left(\frac{m^{2}+pz(z-1)}{4\pi\mu^{2}}\right)^{-s}\right.\notag\\
& \ \ \ \left.+\left(\frac{4\pi \mu^{2}}{R^{-2}}\right)^{s}\sum_{l=1}^{n}\binom{n}{l}E^{c(p,z)^{2}}_{l}(s)\Gm(s)\right]\ .
\end{align}
Here $s\to -1$ and $ s\to 0$ correspond to the physical spacetime dimension $D\to 4$ in\reff{2PVFReg} and\reff{4SVFDR}, respectively. In addition $\sum_{\lbrace p\rbrace}$ denotes the sum over the Maldestam variables. In both cases the first term within the square brackets corresponds to the usual UV divergence present in the analysis of the theory $\mc{L}^{(\unl{0})}$ (Eq.~\eqref{L0}) and the second term is the contribution due to the infinite number of interactions between the zeroth mode and the tower of excited KK modes. Interestingly enough, in these equations is the presence of the inhomogeneous Epstein function\citte{eps03, Ep1992}
\begin{equation}\label{inEps}
E_l^{\al^2}(s)=\sum_{m_1,\cdots ,m_l=1}^{\infty}\frac{1}{\left(m_1^2+\cdots +m_l^2+\al^2\right)^{s}}\ .
\end{equation}
where $\al^{2}$ is either $c^{2}:=m^{2}/R^{-2}$ or $c(p,z)^{2}:=(m^{2}+pz(z-1))/R^{-2}$ depending whether one is considering\reff{2PVFReg} or\reff{4SVFDR}. The Epstein function is a generalization of the well-known Riemann zeta function. Should the UV divergences be susceptible to extraction from these expressions, one may hence well be looking for the corresponding poles at $D\to 4$ and those terms that violate the decoupling theorem \cite{App75}; this will be a plausible hint to renormalize the theory which will be reported in a further communication. 
\vspace{-.4cm}
\section{Acknowledgments}
We acknowledge financial support from CONACyT (Mexico). M.A.L.-O, E.M.-P. and J.J.T. acknowledge SNI (Mexico). M.A.L.-O and E.M.-P  acknowledge PRODEP-SEP (Mexico) as well.

%-------------------------------------------------------------------------------
%%-------------------------------------------------------------------------------
%

\end{document}